\documentclass[
aps,
physrev,
amsmath,
amssymb,
reprint,
]{revtex4-1}

\usepackage[utf8]{inputenc}
\usepackage{graphicx}
\usepackage[caption=false]{subfig}
\usepackage{braket}
\usepackage{natbib}
\usepackage{color}
\usepackage{xcolor}

\newcommand{\tr}{\mathrm{tr}\:}
\newcommand{\im}{\mathrm{Im}\:}

\newcommand{\eh}{E_\mathrm{h}}

\newcommand{\remove}[1]{}

\newcommand{\vi}[1]{{}}
\newcommand{\vii}[1]{{#1}}

\begin{document}

\title{Efficient Compression Of The Environment Of An Open Quantum System}
\author{Max~Nusspickel}
\email{max.nusspickel@kcl.ac.uk}
\author{George~H.~Booth}
\email{george.booth@kcl.ac.uk}

\affiliation{
Department of Physics,
King's College London,
Strand, London, WC2R 2LS, U.K.}

\date{\today}

\begin{abstract}
In order to simulate open quantum systems, many approaches (such as Hamiltonian-based solvers in dynamical mean-field theory) aim for a reproduction of a desired environment spectral density in terms of a discrete set of bath states, mimicking the open system as a larger closed problem.
%
Existing strategies to find a compressed representation of the environment for this purpose can be numerically demanding, or lack the compactness and systematic improvability required for an accurate description of the system propagator.
%
We propose a method in which bath orbitals are constructed explicitly by an algebraic construction based on the
Schmidt-decomposition of response wave functions, efficiently and systematically compressing the description of the full environment.
These resulting bath orbitals are designed to directly reproduce the system Green's function, not hybridization, which allows for consideration of the relevant system energy scales to optimally model.
%
This results in an accurate and efficient truncation of the environment, with applications in a wide range of numerical simulations of open quantum systems.
\end{abstract}

\maketitle

\section{Introduction}
In different fields of physics one is often confronted with the task of describing an open quantum system, coupled to a large environment. This environment can induce dissipation and relaxation effects on the system to (sometimes drastically) change its properties, such as adding finite lifetimes and changing the effective masses of the quasiparticles of the system\cite{RevModPhys.89.015001}.
There are many examples where rather than explicitly treating the open quantum system, the aim is to compress and approximate the environmental effects, such that a compact, closed problem comprising the system coupled to a small set of `bath' states can well approximate the dynamics of the open system of interest. Commonly, this is done such that accurate wave function-based techniques applicable to closed systems can be used, potentially augmented with additional system interactions. Due to the computational scaling of these calculations with respect to the size of this bath space, an efficient protocol for a faithful compression of the environmental effects in these bath states is highly desirable, and is the focus of this work.

The applications of such an approach are numerous. As an example, molecular scale electronics are concerned with junctions where a single molecule is bound to conducting leads, which apply a bias voltage and result in a complicated non-equilibrium many-body problem coupled to many degrees of freedom in the leads\cite{Thoss18,PhysRevB.99.245134}. Molecules with strong effective Coulomb interactions can give rise to emergent phenomena in the system such as Coulomb blockade or Kondo physics\cite{PhysRevLett.103.197202,Parks1370,PhysRevB.83.245415,Park02,Esat16}, while strong vibrational coupling can result in a pronounced change in conductivity and decoherence\cite{Leturcq09,PhysRevB.94.201407}. These are generally treated with wave function approaches such as multi-configurational time-dependent Hartree approximations, and require an efficiently constructed and systematically truncated representation of the leads in order to be effective\cite{Wang15,Wang16}.

In the context of equilibrium many-body physics, quantum embedding methods such as dynamical mean-field theory (DMFT)
\cite{PhysRevLett.62.324,PhysRevLett.70.1666,Georges1996,Sun16} utilize a mapping of the propagator of a macroscopic system to
a smaller correlated open quantum fragment (the so-called ``impurity''), coupled to its environment consisting of the rest of the system. 
In order to solve for the properties of the impurity model with a Hamiltonian-based solver, the large
\vi{or even infinite-sized }
environment needs to be reduced to a tractable number of bath states.
Since \textit{tractable} often means less than 10 bath states for exact-diagonalization (ED) solvers\cite{Liebsch_2011}, or up to 200 for approximate solvers, like density-matrix renormalization group \cite{PhysRevB.90.115124,PhysRevX.5.041032}, selected configuration interaction approaches \cite{PhysRevB.86.165128,PhysRevB.100.125165,PhysRevB.96.085139}, or coupled-cluster singles doubles (CCSD) \cite{PhysRevB.100.115154,Shee19}, an efficient construction of compact bath orbitals faithfully representing the effect of the environment coupling is essential for these methods.

Traditional approaches for the compression of an environment focus on optimally representing its coupling to the open system, as represented by the environment spectral density. While an accurate reproduction of this quantity from the bath states is sufficient to ensure that single-particle dynamics of the system are reproduced, it is not necessarily the most efficient approach, as the quantities of interest ultimately derive from the properties of the system, rather than a faithful representation of this coupling. With this in mind, it is worth attempting to incorporate details of the system Hamiltonian in order to construct a more effective bath space which specifically reproduces the system properties. In this work we demonstrate the analytic construction of a compact bath space which directly reproduces desired properties of the system under the influence of the entire environment, rather than necessarily targeting a faithful representation of the system-environment coupling. This may result in a loss of accuracy in the description of the environment spectral density (the hybridization), but results in a faster convergence of the system properties with respect to bath size, and in an efficient and systematically improvable expansion of the environmental contribution to the system properties.

To achieve this, we use a technique based on the Schmidt decomposition of quantum information theory in order to construct bath orbitals which are represented as contractions over the full set of explicit environmental states.
The method ensures that these bath orbitals exactly span the minimal space necessary to represent a chosen wave function across the full system. By changing the wave functions that we span with the bath orbitals, we can change the expectation values which we target to be conserved in the system space by construction.
In order for these bath states to be represented efficiently as single-particle orbitals, a constraint on the approach is that the properties which are matched must be derived from arbitrary functions of a quadratic system Hamiltonian, in the absence of explicit two-body interactions. This necessitates a mean-field-like approximation to the system Hamiltonian in order to construct approximate bath states. Alternatively, if the system interactions are very strong and substantially change the spectral width of the system propagator, then this modification to the system one-particle properties can be included via the linear coupling to additional fictitious auxiliary terms representing the iterative effect of a system self-energy\cite{Fertitta2018,Fertitta2019,Backhouse20}, as is performed in the iterations of DMFT.

This Schmidt decomposition approach to construct bath orbitals has been used before, in the (energy-weighted) density-matrix embedding theory \cite{PhysRevLett.109.186404,doi:10.1021/ct301044e,Fertitta2018,Fertitta2019} and in $\omega$-DMFT \cite{PhysRevB.91.155107,PhysRevB.101.045126},
where the bath states were selected to reproduce (energy~weighted) density matrices of the system, or the system single-particle propagator (Green's function) and its derivatives at a given frequency, respectively. Similar decomposition methods have also been used elsewhere, for example in order to automatically partition orbital spaces\cite{doi:10.1021/acs.jctc.8b01112}.
The bath orbital construction introduced in this paper aim to generalize these approaches, to represent the dynamics of the single-particle propagator of the system either on the real- or the imaginary-frequency axis more efficiently.
We compare their performance with the traditional approaches of `direct discretization' of the environmental coupling detailed in Ref.~\onlinecite{PhysRevB.92.155126}, and the numerical optimization of bath states\cite{Lin20},
both widely used in DMFT as well as other applications\cite{Liebsch_2011,Zhu20}.
We first briefly introduce the direct discretization method (Sec.~\ref{sec:direct_disc})
and the numerical optimization approach (Sec.~\ref{sec:fit}).
After this, we describe the projection method based on Schmidt decomposition (Sec.~\ref{sec:projection}) and introduce a new way to truncate the bath space for efficient description of equilibrium system dynamics, either for the real frequency or imaginary frequency system propagator.
Application of these bath states to describe the system propagator of the 3d shell of an iron atom in the iron porphyrin molecule, as well as the case of the orbitals of the $\pi$-system of benzene coupled to a continuous hybridization (mimicking the conduction band of a metal surface) will be shown in Sec.~\ref{sec:results}, comparing extensively to direct discretization of the hybridization, as well as numerical fitting of bath states.
\vii{
Additionally, using the example of the iron porphyrin molecule, we test if these results transfer to problems involving explicit two-body interaction terms in the Hamiltonian, and compare the occupation of the 3d states in the presence of different bath orbitals as a compressed environment description at the level of coupled-cluster theory. 
}

\section{Theory}
The general set up for the problem consists of $N_{\rm sys}$ degrees of freedom in a system, coupled to an environment, so that the full Hamiltonian can be written as
\begin{equation}
    H = H_{\rm sys} + H_{\rm env} + H_{\rm coup} .
    \label{eq:ham-sys-env}
\end{equation}
We assume that the environment Hamiltonian is quadratic 
\begin{equation}
    H_{\rm env} = \sum_x^{N_{\rm env}} \epsilon_x c_x^{\dagger} c_x ,
    \label{eq:env}
\end{equation}
and that the coupling to the system is linear in the system operators, as
\begin{equation}
    H_{\rm coup} = \sum_a^{N_{\rm sys}} \sum_{x}^{N_{\rm env}} V_{ax} c^{\dagger}_x c_a + {\rm h.c.} 
    \label{eq:ham_coup}
\end{equation}
Throughout this paper we will index the degrees of freedom of the system with $a,b$ and of the environment with $x,y$.
If the environment contains non-diagonal terms, these can be effectively removed via a diagonalization of $H_{\rm env}$ and rotation of the $c_x^{(\dagger)}$ operators.
\vi{
The number of these environment states ($N_{\rm env}$) can be arbitrarily large, approaching a continuum for the environment spectral density,}
\vii{
While the number of these environment states ($N_{\rm env}$) can be arbitrarily large in principle, and thus resulting in a continuous environment spectral density,}
\
given by
\begin{equation}
    J_{ab}(\omega) = \sum_x^{N_\mathrm{env}} V_{ax}V^*_{bx} \: \delta(\omega - \epsilon_x)
    ,
    \label{eq:j-disc}
\end{equation}
\vii{the projective approach presented in this work requires a finite-sized environment in practice.
We note that a finite-sized approximation to an infinite environment can still be achieved by means of the direct discretization, after which the projective methods discussed later becomes applicable again.
}
We now describe the different approaches to compress the description of the environment in order to reduce the dimensionality of~$H$.

\subsection{Direct discretization}\label{sec:direct_disc}
%
%
%
%
%
%
In the direct discretization method, the environment spectral density (the hybridization spectrum) is directly approximated as a real frequency valued quantity. This is achieved by dividing $\mathbf{J}(\omega)$ (the matrix of the elements of Eq.~\eqref{eq:j-disc}) into $N$ disjoint frequency intervals
and in each interval the spectral density is approximated by a simple set of poles centered at some energy within the interval.
The number of poles in each set is determined by the number of coupling vectors required to reproduce the spectral density at that energy as an outer product,
which is smaller or equal to the number of system orbitals.
The work of de Vega et. al. (Ref.~\onlinecite{PhysRevB.92.155126}) proposes to define energy intervals by the requirement that each interval $[\omega_n, \omega_{n+1}]$ should have the same ratio of the total environment spectral weight
\begin{equation}
    \int_{\omega_n}^{\omega_{n+1}} \tr \mathbf{J}(\omega) \: \mathrm{d}\omega = \frac{1}{N} \int_{\omega_\mathrm{min}}^{\omega_\mathrm{max}} \tr \mathbf{J}(\omega) \: \mathrm{d}\omega
    ,
    \label{eq:equal-spectral-weight}
\end{equation}
with $\omega_0 = \omega_\mathrm{min}$ and $\omega_{N-1} = \omega_\mathrm{max}$.
Different applications may require a different choice of intervals, for example in the numerical renormalization group method a logarithmic discretization with more discretization points towards the low-energy region is standard
\cite{RevModPhys.80.395}.
In this work, however, we only consider the division according to equal environment spectral weight of each interval.
Note that Ref.~\onlinecite{PhysRevB.92.155126}, where this equal spectral weight division was proposed, only considers the case of a single degree of freedom in the system, i.e., the spectral density is a scalar function, instead of a matrix function.
We generalized this in Eq.~\eqref{eq:equal-spectral-weight} by including the trace operator.

For each interval, the set of coupling vectors which produce the integrated spectral density can be obtained by diagonalization
\vi{
\begin{equation*}
    \int_{\omega_n}^{\omega_{n+1}} J_{ab}(\omega) \: \mathrm{d}\omega
    = \sum_{y}^{N_\mathrm{sys}} \tilde{V}_{a y} \lambda_{y} \tilde{V}_{b y}
    = \sum_{y}^{N_\mathrm{sys}} V_{a y} V_{b y}
    ,
\end{equation*}
with $V_{a y} = \sqrt{\lambda_x} \tilde{V}_{a y}$.
Note that $\mathbf{J}(\omega)$ is positive semidefinite due to the outer product form of Eq.~\eqref{eq:j-disc}.
}

\vii{
\begin{equation}
    \int_{\omega_n}^{\omega_{n+1}} J_{ab}(\omega) \: \mathrm{d}\omega
    = \sum_{x}^{N_\mathrm{sys}} \tilde{V}_{a x}^{(n)} \lambda_{x}^{(n)} \tilde{V}_{b x}^{(n)}
    = \sum_{x}^{N_\mathrm{sys}} V_{a x}^{(n)} V_{b x}^{(n)}
    \label{eq:direct_diag}
    ,
\end{equation}
where $\tilde{V}^{(n)}$ and $\lambda^{(n)}$ are the eigenvectors and eigenvalues of the integrated spectral density in the interval $[\omega_n,\omega_{n+1}]$ 
and $V_{a x}^{(n)} = \sqrt{\lambda_x^{(n)}} \tilde{V}_{a x}^{(n)}$.
Note that $\mathbf{J}(\omega)$ is positive semidefinite at all $\omega$ due to the outer product form of Eq.~\eqref{eq:j-disc} and thus all $\lambda_x^{(n)} \ge 0$
Finally, taking the union of the couplings $V_{ax}^{(n)}$ over all intervals $n$ yields an approximate set of system-environment couplings $V_{ax} = \cup_{n} V_{ax}^{(n)}$ over the interval $[\omega_\mathrm{min},\omega_\mathrm{max}]$. 
}
This diagonalization step replaces the simpler operation of taking the square-root in case of a scalar $\mathbf{J}(\omega)$, but leads to the addition of up to ${N_\mathrm{sys}}$ bath states for each interval.
Finally, the corresponding bath state energies $\epsilon_n$ are calculated as interval averages
\begin{equation}
    \epsilon_n = \frac{\int_{\omega_n}^{\omega_{n+1}} \omega \: \tr \mathbf{J}(\omega) \: \mathrm{d}\omega}
    {\int_{\omega_n}^{\omega_{n+1}} \tr \mathbf{J}(\omega) \: \mathrm{d}\omega}
    \label{eq:direct_avg_energy}
    .
\end{equation}
We note that the direct discretization method can be regarded as a pole merging procedure:
In each frequency interval all poles are shrunk into a single energy and the corresponding combined spectral density
is decomposed into a set of $N_\mathrm{sys}$ coupling vectors.
In this way, the approach is systematically improvable to the exact description of the environment spectral density as the number of intervals increases, correspondingly increasing the number of effective bath states representing this environment.
An alternative pole merging procedure for a scalar $\mathbf{J}(\omega)$ was described in Ref.~\onlinecite{PhysRevB.90.085102} and used in a real frequency DMFT formalism.

In some applications the exact environment spectral density is not explicitly known.
An example for this is DMFT after the introduction of a self-energy in the environment. If this self-energy can be expressed in an explicit pole (Lehmann) representation, then the environment Hamiltonian can again can expressed in the form of Eq.~\ref{eq:env} and the direct discretization method used as above. If only the hybridization function 
\begin{equation}
    \Delta_{ab}(z) = \int \frac{J_{ab}(\omega)}{z - \omega} \: \mathrm{d}\omega
    = \sum_{x}^{N_\mathrm{env}} \frac{V_{ax} V_{bx}}{z - \epsilon_x}
    \label{eq:hybrid_poles}
\end{equation}
is known, then it can be related to the environment spectral density via the Sokhotski-Plemelj theorem
\begin{equation}
    \mathbf{J}(\omega) = \lim_{\eta \to 0^+} -\frac{1}{\pi} \mathrm{Im} \: \mathbf{\Delta}(\omega + \mathrm{i} \eta)
    \label{eq:j-sokhotski}
    ,
\end{equation}
with the caveat that any finite broadening ($\eta$) will lead to an error compared to the exact spectral density.
However, in other cases the hybridization is only known on the Matsubara axis far from the real frequency axis. 
In these cases, analytic continuation could be performed to obtain the spectral density, although the smoothing of the spectral features in this will again result in further approximation.
On the other hand, if DMFT is performed on the real axis with a sufficiently small broadening $\eta$,
according to Eq.~\eqref{eq:j-sokhotski}, the scaled imaginary part of the hybridization could be used as an (broadened) approximation of the exact spectral density\cite{Zhu20}.

\subsection{Numerical optimization}\label{sec:fit}
Often it is sufficient to find a discretization of the bath, which approximately reproduces the hybridization along the Matsubara axis,
without the need for a good match along the real axis.
This is often performed for DMFT applications with Hamiltonian-based solvers\cite{caffarel1994exact,Liebsch_2011,Lin20}, where a distance functional of the form
\begin{equation}
    d[\mathbf{\Delta}_\mathrm{disc}] = \sum_{n=0}^N w(\omega_n) \mathrm{tr} \: |\mathbf{\Delta}(\mathrm{i}\omega_n) - \mathbf{\Delta}_\mathrm{disc}(\mathrm{i}\omega_n) |^2 \label{eq:numcostfn}
\end{equation}
is minimized, where $\mathbf{\Delta}_\mathrm{disc}$ is the discretized hybridization,
\begin{equation}
    \left[ \mathbf{\Delta}_\mathrm{disc} (z) \right]_{ab} = \sum_k^{N_{\rm bath}} \frac{V_{ak} V_{bk}^*}{z - \epsilon_k}
    \label{eq:hybrid_disc},
\end{equation}
and $\{V_{ak}, \epsilon_k \}$ denotes the system-bath coupling and bath energy for state $k$. The frequency weighting function $w(\omega_n)$ controls the desired focus of the fit onto high or low energy Matsubara dynamics (taken in this work to be $1/\omega_n$), while $N$ determines how many Matsubara points are included in the fit.

\vi{
The numerical minimization of Eq.~\ref{eq:numcostfn} works, since the hybridization (and $d[\mathbf{\Delta}_\mathrm{disc}]$) is a smooth function on the Matsubara axis
and only weakly dependent on the detailed features of the spectral density, especially at high energies. 
}
\vii{
The numerical minimization of Eq.~\ref{eq:numcostfn} works, since the hybridization is a smooth function on the Matsubara axis and only weakly dependent on the detailed features of the spectral density, especially at high energies.
As a result, the optimization surface defined by the functional $d[\mathbf{\Delta}_\mathrm{disc}]$ also varies smoothly with variations of the fitting 
parameters.
}

However, this weak dependence becomes a disadvantage if one would like to reconstruct the spectral density from the Matsubara hybridization,
as the required analytic continuation is generally ill-conditioned. Furthermore, the numerical minimization of the cost function in Eq.~\ref{eq:numcostfn} is computationally expensive for larger numbers of bath orbitals and system degrees of freedom, with the number of parameters to numerically optimize scaling as $\mathcal{O}[N_{\rm bath} N_{\rm sys}]$.
The robustness of this optimization has also been called into question as the hybridization function becomes more complex with large off-diagonal contributions, ensuring that it is often not a favoured approach (and sometimes even a limiting step) for large impurity and bath spaces in DMFT\cite{PhysRevB.100.125165}. However, recent work using semi-definite programming to automatically enforce the physically allowed structure of Eq.~\ref{eq:hybrid_disc} holds promise to improve the robustness of this approach \cite{Lin20}.
\vii{
Additionally, a regularized distance functional, based on a data-science approach, has been proposed to improve the efficiency of the numerical fitting approach \cite{shinaoka2020sparse}.
}

\section{Projective approach}\label{sec:projection}
The direct discretization and numerical optimization methods define the bath states solely based on the environment Hamiltonian and its coupling to the system, $H_{\rm env}$ and $H_{\rm coup}$. They are devised such that regions of large environment spectral density (hybridization) are most faithfully represented, and to minimize the overall error made in the description of the environment. However, if there are regions where the hybridization is large, but which is well outside the effective energy scales of the system, then these environmental states will correspondingly have a small affect on the system properties, despite a potentially large coupling. This results in bath states which are representing energy scales which are not required for a faithful representation of system dynamics. 

The projection approach here aims to take this into account, by explicitly computing bath orbitals which are required in order to exactly match a particular system property. This builds information from the whole Hamiltonian into the reduced dimensionality environment construction, rather than neglecting the form and spectral range of $H_{\rm sys}$. The constraint for this exact construction in order to ensure single-particle bath states is that the system Hamiltonian is of quadratic form, i.e.,
\begin{equation}
    H_{\rm sys} = \sum_{ab}^{N_{\rm sys}} h_{ab} c^{\dagger}_a c_b . \label{eq:quad_sys}
\end{equation}
In DMFT, this is exactly the form required in the construction of bath states, since the impurity model which aims to be matched consists of the uncorrelated impurity (system) Hamiltonian of quadratic form, connected to the environment represented by the hybridization, which can be cast in the form of Eq.~\ref{eq:hybrid_poles} \cite{Liebsch_2011} \vii{(see Appendix~\ref{ssec:dmft} for a detailed proposal of how this approach would be combined with DMFT)}.
In other applications, an approximate quadratic form of the system Hamiltonian could be formed, potentially including self-energy effects of the system as explicit additional terms in the system Hamiltonian
\vii{ and thus taking electron correlation into account when constructing the bath}
\cite{Fertitta2018,Fertitta2019}.
%

In contrast to the direct method and numerical minimization method which determine fictitious bath parameters to match the environment spectral density, the projection method explicitly constructs a set of bath orbitals that span some subspace of the complete environment. While this bath space is determined by the full Hamiltonian, the parameters of the bath space are then determined via projection of $H_{\rm env}$ and $H_{\rm coup}$ into this subspace.
%
In this way, the bath orbitals generated by the projection method can never be ``wrong'', but only more or less efficient in
their ability to describe the desired influence of the environment. As the number of bath orbitals increases, the space spanned must rigorously approach the full space of the environment. When the number of bath orbitals is equal to $N_{\rm env}$, then the bath space is complete, and any further orbitals will simply be in the linear span of the previous bath orbitals. In this way, the bath space construction is efficient, and trivially systematically improvable to an exact description of the original environmental space.
%

%
%
As discussed in Sec.~\ref{sec:direct_disc}, in some cases the hybridization is only known as a function of (possibly imaginary) frequency and not in a pole or Lehmann representation amenable to rewritting in the form of Eq.~\ref{eq:j-disc}.
In these cases, however, another strategy, like the direct method or numerical minimization with a larger number of discretization intervals or
fit parameters can be used first in order to determine a suitable description of the environment via the energies $\epsilon_x$ and couplings $V_{ax}$,
without the need for a compact representation.
The projection method can then subsequently be applied in order to reduce the environment degrees of freedom to a manageable size of bath degrees.
The quantity conserved over the system degrees of freedom depends on the state chosen to be decomposed, from which the bath orbitals spanning the environmental subspace are defined. Any expectation value can be split into the overlap of two (unnormalized) wave functions. As long as each of these wave functions is decomposed into its appropriate compressed bath space via the Schmidt decomposition, then the bath space required to exactly match the individual wave functions over the system for the expectation value of interest will be spanned.

In this work, the bath orbitals for the projection method are derived from the Schmidt decomposition
of a wave function with the general form
\begin{equation}
    f(H) c_a^{(\dagger)} \ket{\Psi}
    ,
\end{equation}
where $f(H)$ is some function of the full Hamiltonian
given by Eqs.~(\ref{eq:ham-sys-env}--\ref{eq:ham_coup}, \ref{eq:quad_sys})
and $\ket{\Psi}$ the ground state of the latter. This form allows for arbitrary probes of the one-particle physics of the system, via the possible expectation values formed from overlaps of these states, as well as ensuring that the Schmidt decomposition will provide single-particle orbitals.
Denoting the eigenvalues and eigenvectors of $H$ as $\lambda_i$ and $C_{i}$,
the lesser (occupied) and greater (virtual) bath orbitals resulting from the decomposition have the form
\begin{alignat}{2}
\ket{b_a^{\mathrm{<}}(Q)} &=
\sum_{i: \lambda_i < \mu}
\sum_{x}^{N_\mathrm{env}}
C_{ai} C_{xi} f(\lambda_i ; Q)
\ket{x}
\label{eq:bath-general-occ}
\\
\ket{b_a^{\mathrm{>}}(Q)} &=
\sum_{i: \lambda_i > \mu}
\sum_{x}^{N_\mathrm{env}}
C_{ai} C_{xi}
f(\lambda_i; Q)
\ket{x}
\label{eq:bath-general-vir}
,
\end{alignat}
where we use $Q$ to denote some set of parameters which the function $f(\lambda_i;Q)$ may depend on parametrically.
We can define the `power' bath orbitals, introduced for EwDMET\cite{Fertitta2018,Fertitta2019} which employ the power functions
\begin{equation}
    f(\lambda_i; n) = \lambda_i^n \quad n \in \{0, 1, 2,\dots\}
    \label{eq:f_static}
\end{equation}
and which exactly span the space required to match the $n^{\rm th}$-order energy-weighted density matrices, defined as
\begin{align}
T_{ab}^{<,(n)} &= \langle \Psi | c^{\dagger}_b [c_a, H]_{\{ n \}} | \Psi \rangle  \\
T_{ab}^{>,(n)} &= \langle \Psi | [c_a, H]_{\{n\}} c_b^{\dagger} | \Psi \rangle ,
\end{align}
where $[c_a,H]_{\{n\}}=[...[[c_a,H],H],\dots H]$ with $n$ total commutators, and $[c_a,H]_{0} = c_a$. These system observables correspond to the moments of the spectral distribution for the particle and hole propagators of the system \cite{Fertitta2019}, with the zeroth-order moment representing the single-particle density matrix of the system, conserved in the DMET construction \cite{PhysRevLett.109.186404}.
It should be noted that by including all powers of Eq.~\eqref{eq:f_static} individually up to some degree $n_\mathrm{max}$, the space of (all) polynomials of degree $n_\mathrm{max}$ is spanned, regardless of their specific form.
As an example, the power bath orbitals of EwDMET span a space to ensure the ability to reproduce the system spectral functions defined in a basis of Chebyshev polynomials to the same degree\cite{PhysRevB.90.115124}. 
In Ref.~\onlinecite{PhysRevB.91.155107}, `dynamic' bath orbitals were introduced with the function
\begin{equation}
    f(\lambda_i; z) = \frac{1}{z - \lambda_i}
    \label{eq:dynamic_bath}
    ,
\end{equation}
which ensure the matching of the particle and hole Green's functions at the complex frequency~$z$ with $\im(z) \neq 0$.
These were further generalized to provide the derivatives of the Green's function at any frequency in Ref.~\onlinecite{PhysRevB.101.045126}. While these gave compact bath spaces for describing the interacting Green's function of the system, the dynamical nature of the bath introduced additional difficulties.
\vii{
We note that other choices of kernel functions $f$ are possible and in appendix~\ref{ssec:legendre} we discuss one such choice, which would match the Legendre representation of the imaginary time Green's~function.
}

In this work we build on the imaginary part of these dynamic bath orbitals, but ensure a fixed, static bath space.
The reasoning behind this is that the real and imaginary part of Green's~functions and self-energies are not independent, but related via
the Kramers-Kroning relations.
In our previous work of Ref.~\onlinecite{PhysRevB.101.045126} we were concerned with a pointwise matching of the values and derivatives of the Green's function
between a lattice and impurity space.
Since the description of the Green's function away from this frequency point can in principle be poor, matching only the imaginary part of the Green's function
does not guarantee a matching of the real part at the same frequency via the Kramers-Kroning relations, since they contain an integration over the whole frequency axis.
In Ref.~\onlinecite{PhysRevB.101.045126} we thus included real and imaginary part of the dynamic bath orbitals independently.
In this work we aim to reproduce the Green's~function more globally along the frequency axis and thus only use the imaginary part to construct orbitals.

We focus on two choices of frequency points, one where the complex frequencies are located on the shifted real frequency axis, i.e., $z = \omega_m + \mathrm{i}\Lambda$ with $\omega_m$ being uniformly distributed in some frequency interval $[\omega_\mathrm{min}, \omega_\mathrm{max}]$,
and the other where we use the Matsubara points $z = \mathrm{i}\omega_n = \mathrm{i} (2n + 1)\pi/\beta$ with $\beta$ defining an inverse temperature scale.
With these choices of complex frequencies, ensuring the matching of the imaginary part of the system propagator results in a kernel for the bath orbitals as
\begin{equation}
    f(\lambda_i;\omega_m,\Lambda) = - \frac{\Lambda}{(\omega_m - \lambda_i)^2 + \Lambda^2}
    \label{eq:spectral_orbitals}
\end{equation}
for the shifted real frequencies, and
\begin{equation}
    f(\lambda_i;\omega_n) = - \frac{\omega_n}{\lambda_i^2 + \omega_n^2}\
    \label{eq:matsubara_orbitals}
\end{equation}
for the Matsubara frequencies.
Inserting these functions into Eqs.~(\ref{eq:bath-general-occ}, \ref{eq:bath-general-vir}) yields the corresponding bath orbitals representing the environment,
which we term `spectral orbitals' [using Eq.~\eqref{eq:spectral_orbitals}] and `Matsubara orbitals' [using Eq.~\eqref{eq:matsubara_orbitals}], respectively.
Note that for $\Lambda \to \infty$ in Eq.~\eqref{eq:spectral_orbitals} or $\omega_n \to \infty$ in Eq.~\eqref{eq:matsubara_orbitals}, the function becomes a constant and insensitive to the eigenvalues $\lambda_i$.
As a result, in these limiting cases, the kernel function becomes identical to the zeroth-order power orbitals of Eq.~\eqref{eq:f_static}.

In contrast to our previous work, where we selected a small set of evolving (dynamic) bath orbitals for each frequency considered, here we construct spectral or Matsubara bath orbitals over a large set of frequency points.
The orbitals are then orthonormalized by diagonalizing the outer product matrix $\mathbf{P}$ with elements
\begin{equation}
    P_{xy} = \sum_{\zeta \in \{<,>\}} \sum_a^{N_\mathrm{sys}} \sum_Q \braket{x | b_a^\zeta(Q)} \braket{b_a^\zeta(Q)|y}
    \label{eq:outer_product_matrix}
    ,
\end{equation}
where $x,y$ are the eigenstates of Eq.~\eqref{eq:env}.
Only after this orthonormalization do we reduce the number of chosen bath orbitals by selecting the eigenvectors of $\mathbf{P}$ which correspond to the $N_\mathrm{bath}$ largest eigenvalues (due to the outer product form, $\mathbf{P}$ is positive semidefinite). This relates to choosing the largest Schmidt vectors to represent the linear span required to reproduce the system spectral function in each domain.
Note that this procedure requires that the orbitals $\ket{b_a^\zeta(Q)}$ are not normalized before $\mathbf{P}$ is constructed and their normalization simply follows
from Eqs.~(\ref{eq:bath-general-occ}, \ref{eq:bath-general-vir}) and the inserted functions \eqref{eq:spectral_orbitals} or \eqref{eq:matsubara_orbitals}, respectively.
In this way, the norm of the orbitals is large, if the system-environment coupling is strong [via $C_{ai}C_{xi}$ in Eqs.~(\ref{eq:bath-general-occ},\ref{eq:bath-general-vir})] and/or the energies of the environment are close the real frequency of the orbital [for spectral orbitals via Eq.~\eqref{eq:spectral_orbitals}] or the absolute energies are close to the magnitude of the imaginary frequency [for Matsubara orbitals via \eqref{eq:matsubara_orbitals}].

As a result, the largest eigenvectors of $\mathbf{P}$ represent linear combinations of environment orbitals with large norm, i.e., those which are most significant in describing the overall effect of the environment on the real frequency (spectral orbitals) or imaginary frequency (Matsubara orbitals) Green's~function, respectively.
We note that as an result of this truncation in the eigenspace of $\mathbf{P}$, the property of exact conservation of the Green's function at the chosen set of frequencies is lost.
Instead, the $N_\mathrm{bath}$ selected eigenvectors efficiently approximate the Green's function across the selected frequency range, with a well-defined and efficient truncation criteria which can be systematically improved.
\vi{
If additionally the exact conservation of a property is desired, for example of the density matrix via the zero-order power orbitals, one can add the corresponding orbitals to the bath space, and then orthonormalize them with respect to the selected eigenvectors of $\mathbf{P}$.
}
\vii{
If additionally the exact conservation of a property is desired, for example of the density matrix via the zero-order power orbitals, one can construct these first, remove their linear span from the environment space $x$ in Eqs.~(\ref{eq:bath-general-occ},\ref{eq:bath-general-vir}), and add them to the selected eigenvectors of $\mathbf{P}$ to obtain the final set of bath orbitals.
}

Once a set of $N_\mathrm{bath}$ orthonormal bath states $\{\ket{\beta}\}$ has been constructed this way, the Hamiltonian representing the system and compressed environment is found by projecting the full Hamiltonian of Eq.~\eqref{eq:ham-sys-env} into the system $\oplus$ bath space via
\begin{equation}
    H' = R^\dagger H R
    \label{eq:projection-h}
    ,
\end{equation}
with
%
%
\begin{equation}
    R = R_\mathrm{sys} + R_\mathrm{bath} = \sum_a^{N_\mathrm{sys}} \ket{a}\bra{a} +  \sum_\beta^{N_\mathrm{bath}} \ket{\beta}\bra{\beta}
    .
\end{equation}  
The approximate hybridization, which $H'$ represents, is then given by the right term of Eq.~\eqref{eq:hybrid_poles}, with the poles $\epsilon_x$ being the eigenvalues of $R_\mathrm{bath}^\dagger H' R_\mathrm{bath}$ and the couplings $V_{x}$ being $R_\mathrm{sys}^\dagger H' R_\mathrm{bath} U$, where the columns of $U$ are the eigenvectors of $R_\mathrm{bath}^\dagger H' R_\mathrm{bath}$ (which rotate the bath orbitals into its eigenbasis). 

\vi{
As a final note, we remark that the orbitals of Eqs.~(\ref{eq:bath-general-occ},~\ref{eq:bath-general-vir}) can be used to span other wave functions depending on the choice for the function $f(\lambda_i)$.
For example, it is well known that imaginary time Green's function can be represented efficiently in terms of orthogonal polynomials like the Legendre polynomials \cite{PhysRevB.84.075145,doi:10.1021/acs.jctc.5b00884,doi:10.1021/acs.jctc.5b00884,PhysRevB.98.075127}.
To construct bath orbitals which span the space which can fully describe the Green's function in the Legendre representation of a given degree $l$, one can simply use
\begin{equation}
    f(\lambda_i;l) = \int_{-\beta}^0 P_l[x(\tau)] G_{\lambda_i}(\tau) \: \mathrm{d} \tau
    ,
\end{equation}
where $P_l$ is the Legendre~polynomial of degree $l$, $x(\tau) = 2\tau / \beta + 1$
maps the interval of integration~$[-\beta, 0]$ to the interval~$[-1, 1]$, and
$G_{\lambda_i}(\tau)$ is the imaginary time Green's function
\begin{equation}
    G_{\lambda_i}(\tau) =
    \frac{\mathrm{e}^{-\lambda_i \tau}}{1 + \mathrm{e}^{\lambda_i \beta}}
    .
\end{equation}
Including these bath orbitals up to an order $l_\mathrm{max}$ results in an efficient representation of the imaginary time and frequency dynamics, but in practice we find that the Matsubara orbitals with subsequent truncation of $\mathbf{P}$ are even more compact.
}

\section{Results}
\label{sec:results}
We test the performance of the spectral and Matsubara bath orbitals introduced in Sec.~\ref{sec:projection} on two contrasting cases of open systems with a quadratic  Hamiltonian, to see how faithfully and compactly they can represent the effect of their environment.
These are:
\begin{enumerate}
\item The 3d atomic shell of the iron atom of an iron porphyrin molecule ($\mathrm{FeN_4C_{20}H_{12}}$) described by a mean-field Hamiltonian.
\item The $\pi$-orbitals of a benzene molecule ($\mathrm{C_6H_6}$) coupled to a continuous, model hybridization of a metallic conduction band.
\end{enumerate}
In the first case, the poles of the environment spectral density are directly known, whereas in the second case the hybridization is only known as a parameterized  function, taken to be constant within some interval, commonly used as a model of leads in molecular junctions.

\subsection{3d Shell of Iron Porphyrin}
Due to its biological significance, the iron porphyrin molecule has been studied extensively, with many different computational methods\cite{doi:10.1021/acs.jctc.6b00714,doi:10.1021/ct500103h,doi:10.1021/jp0441442}.
Electrons in the partially filled 3d atomic orbitals of the central iron atom experience strong electron--electron interactions, which constitute a difficult problem for density functional theory and may necessitate a multireference treatment for high accuracy results\cite{PANCHMATIA200847,doi:10.1021/ct900567c}.
%
In this context, quantum embedding methods can be an useful alternative, since they are more flexible in the choice of the correlated (``impurity'') subspace, which can be made up of partially occupied, local degrees of freedom, than quantum chemical multireference methods\cite{PhysRevLett.110.106402}.
Since quantum embedding methods often utilize computationally demanding, high accuracy quantum impurity solvers, an efficient description of the environment is crucial, which makes iron porphyrin an ideal test case for the spectral and Matsubara bath orbitals.

We first perform a restricted Hartree--Fock (RHF) calculation using the cc-pVTZ basis set \cite{doi:10.1063/1.456153} (956 orbitals) and determine intrinsic atomic orbitals~\cite{Knizia2013} (IAOs), using the quantum chemistry framework PySCF \cite{doi:10.1002/wcms.1340}.
Five of the constructed IAOs have large overlap with the five 3d basis functions centered at the iron atom
and we use these as the system (the ``impurity''), with the remaining IAOs plus the eigenvectors of the projector into
the remaining space (orthogonal complement to the IAO space) defining the environment orbitals.
Taken all together, these states define a rotation of the original canonical Hartree--Fock orbitals.
%
%
The exact hybridization (on the level of RHF), can be calculated as discussed in Sec.~\ref{sec:projection}, with the Hamiltonian $H'$ in Eq.~\eqref{eq:projection-h} being the full Fock matrix of the RHF calculation (i.e., $R$ is set to unity).
\vii{
The imaginary part of the hybridized Green's function of the 3d orbitals, as well as the corresponding hybridization are shown in Fig.~\ref{fig:porphyrin-spectrum}.
\begin{figure}
    \centering
    \includegraphics[width=\linewidth]{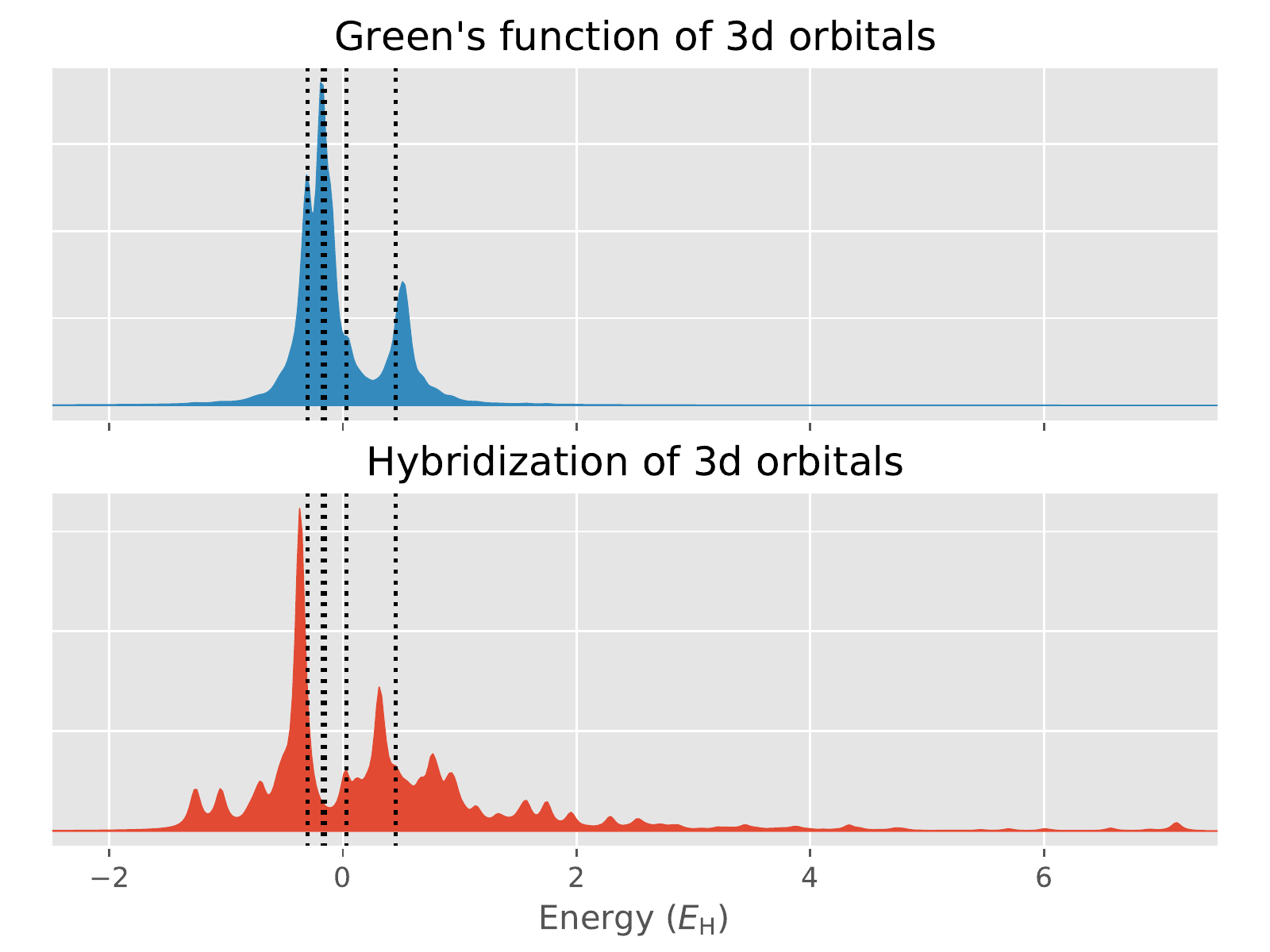}
    \caption{Imaginary parts of the hybridized Green's function (top row) of the 3d states of iron porphyrin and the imaginary part of the corresponding hybdridization function (bottom row). The vertical dotted lines show the energies of the unhybridized 3d states.}
    \label{fig:porphyrin-spectrum}
\end{figure}
From this figure alone, one can already see that due to the large environment the hybridization function can exhibit a complex structure and have spectral weight over a large energy range.
In fact, the core states of the molecule lead to features in the hybridization far below the energy range shown in Fig.~\ref{fig:porphyrin-spectrum} (for example the 1s~orbital of iron lies at $\approx -261~E_\mathrm{H}$).
On the other hand, the hybridized system's Green's~function shows far less complexity.
It is thus easy to see that a matching of the Green's~function will be more simply achieved than matching the hybridization.
}

\vi{
We then construct the spectral orbitals via Eqs.~\eqref{eq:spectral_orbitals} and (\ref{eq:bath-general-occ}, \ref{eq:bath-general-vir}), using 1000 uniform frequency points in the interval between
$\omega_\mathrm{min} = -10~\eh$ and $\omega_\mathrm{max} = 10~\eh$ and with a broadening of $\Lambda = 0.05~\eh$.
}
\vii{
Following the Hartree--Fock calculation, we construct the spectral orbitals via Eqs.~\eqref{eq:spectral_orbitals} and (\ref{eq:bath-general-occ}, \ref{eq:bath-general-vir}), using 1000 uniform frequency points in the interval between
$\omega_\mathrm{min} = -10~\eh$ and $\omega_\mathrm{max} = 10~\eh$ and with a broadening of $\Lambda = 0.05~\eh$.
}
These states are orthonormalized by diagonalizing the outer product matrix of Eq.~\eqref{eq:outer_product_matrix} and selection of eigenvectors with the largest eigenvalues as discussed in Sec.~\ref{sec:projection}.
We compare this to the direct discretization method using the equal spectral weight criterion of Eq.~\eqref{eq:equal-spectral-weight} in the same interval $[-10~\eh, 10~\eh]$.
Note that we use a broadened ($\eta = 0.05~\eh$) imaginary part of the hybridization instead of the spectral density $\mathbf{J}(\omega)$ for the determination of the intervals according to Eq.~\eqref{eq:equal-spectral-weight}.
Once these intervals are found, however, we use the unbroadened spectral density to calculate the couplings and average energies according to Eqs.~(\ref{eq:direct_diag},~\ref{eq:direct_avg_energy}).

\begin{figure}[!htb]
    \centering
    \includegraphics[width=\linewidth]{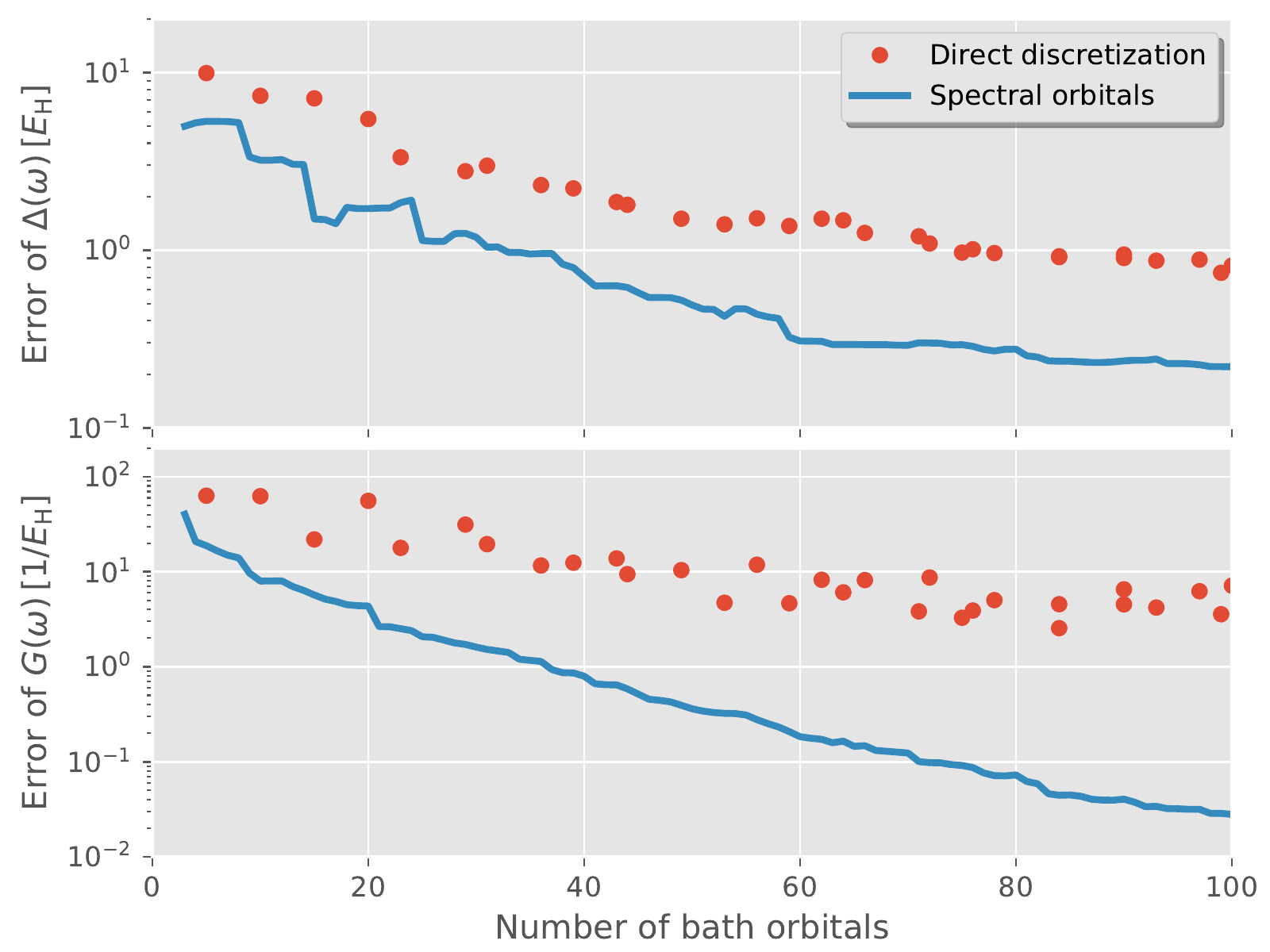}
    \caption{
    Error of the retarded (real-frequency) hybridization (top row) and Green's function (bottom row)
    defined according to Eq.~\eqref{eq:error-ret} for the Fe 3d shell of iron porphyrin.
     The `Direct discretization' refers to the method of Sec.~\ref{sec:direct_disc}, and the `Spectral orbitals' to the projected bath orbitals with the kernel of Eq.~\eqref{eq:spectral_orbitals}.
    Since the direct discretization method is very sensitive to the intervals defined by the equal spectral weight criterion, the decay of the errors (especially of the Green's function) is noisy and not monotonic.
    }
    \label{fig:porphyrin-ret}
\end{figure}

To compare the errors of both methods, we calculate the modelled hybridization function of each bath space, $\Delta'(\omega)$, as well as resulting Green's~function
\begin{equation}
    G'(\omega) = \frac{1}{\omega - H_\mathrm{sys} - \Delta'(\omega)} 
\end{equation}
and compare them to the exact quantities $\Delta(\omega)$ and $G(\omega)$ with the full environment, using the metric
\begin{equation}
    \chi[A] = \sqrt{ \sum_{mab} |A_{ab}(\omega_m + \mathrm{i}\eta) - A'_{ab}(\omega_m + \mathrm{i}\eta)|^2}
    ,
    \label{eq:error-ret}
\end{equation}
where $A$ can be either $\Delta$ or $G$, depending on whether the hybridization or system Green's function is to be compared.  
For the evaluation of Eq.~\eqref{eq:error-ret}, we use the same uniform frequency points in the interval $[-10~\eh,10~\eh]$ and the same broadening $\eta = \Lambda = 0.05~\eh$, as for the spectral orbitals.
We note that defining a fair metric for comparison of the accuracy of real-frequency spectral functions is difficult. However, while the quantitative errors presented will depend on the choice of $\eta$, the qualitative results hold for a range of this parameter, and the general conclusions are sound.
Figure~\ref{fig:porphyrin-ret} shows the error $\chi$ of the bath orbitals from spectral projection and direct discretization in the real-frequency domain, for different numbers of bath orbitals.
In this systems the spectral orbital method performs better than the direct discretization method,
both with respect to the error of the hybridization and of the Green's function.
20 spectral orbitals have a similar accuracy as 80 direct bath states when measuring the error of the Green's function.
Furthermore, the efficiency of the spectral orbitals improves even more with larger number of bath orbitals. It should be noted that for the direct discretization, the error in the hybridization is relatively monotonically decreasing, while the errors in the described system Green's function are larger and less systematic. In contrast, the spectral orbitals give a generally better and more systematic description of $G(\omega)$ than $\Delta(\omega)$, reflecting their design approach. Nevertheless, the orbitals still give a better description of both quantities compared to the direct discretization method.
\begin{figure}[!ht]
    \centering
    \includegraphics[width=\linewidth]{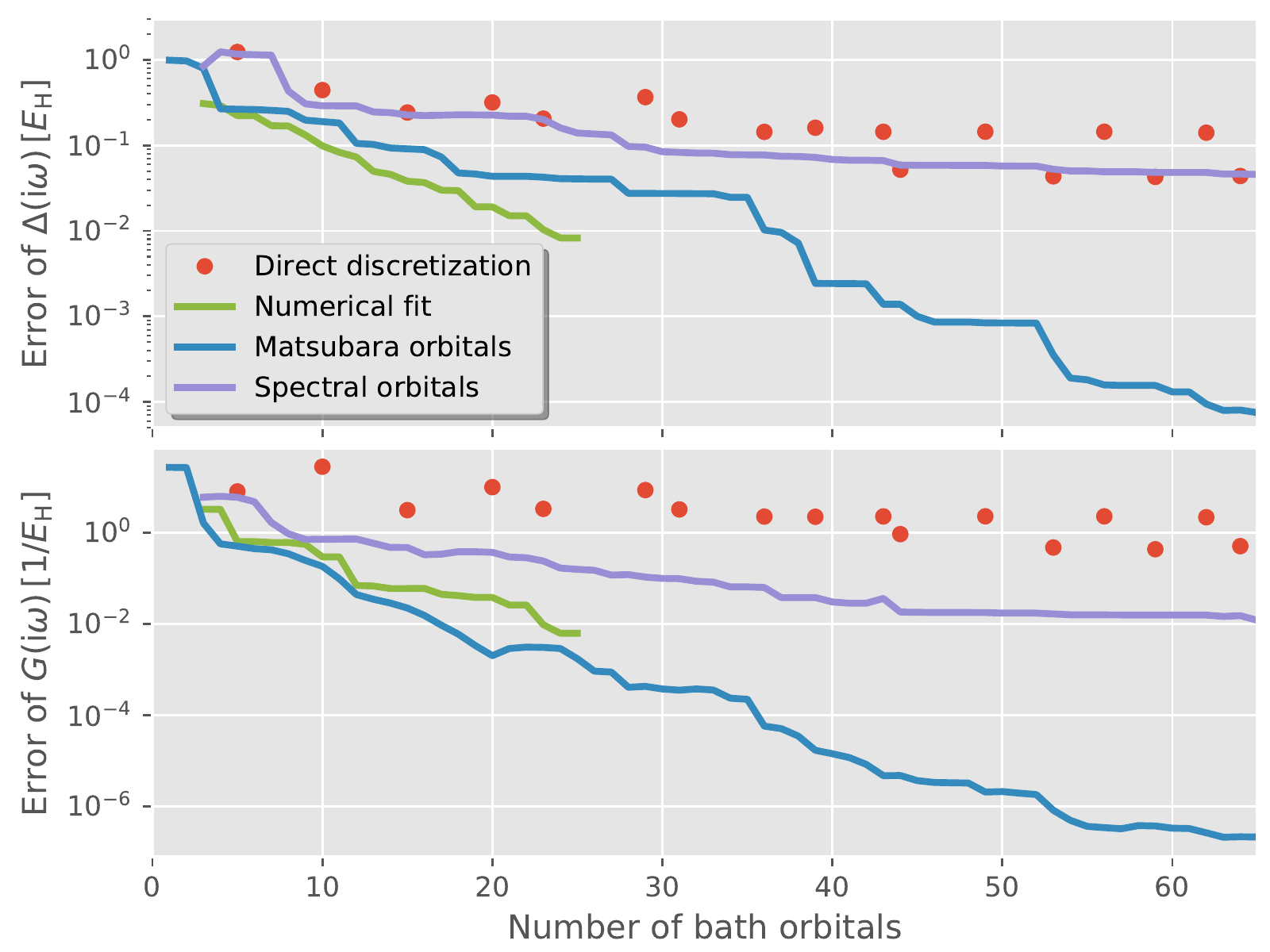}
    \caption{
    Error of the Matsubara hybridization (top row) and system Green's function (bottom row)
    defined according to Eq.~\eqref{eq:error-msb} for the Fe 3d shell of iron prophyrin. The `Numerical fit' refers to the method in Sec.~\ref{sec:fit}, while the `Matsubara orbitals' refer to the projected bath orbitals with the kernel of Eq.~\eqref{eq:matsubara_orbitals}.
    }
    \label{fig:porphyrin-msb}
\end{figure}
Next we test the Matsubara orbitals, which are designed to optimally reproduce the Matsubara Green's~function, for the same system.
We calculate the Matsubara hybridization and Green's~function on the first 1000 Matsubara points with the inverse temperature $\beta = 100~\eh^{-1}$.
Similar to Eq.~\eqref{eq:error-ret}, we measure the error of the Matsubara system Green's function and hybridization as
\begin{equation}
    \chi[A] = \sqrt{ \sum_{nab} |A_{ab}(\mathrm{i}\omega_n) - A'_{ab}(\mathrm{i}\omega_n)|^2}
    .
    \label{eq:error-msb}
\end{equation}
In addition to the direct discretization and Matsubara orbital methods, we also compare to a direct numerical optimization of the bath orbitals as described in Sec.~\ref{sec:fit}. These `numerical fit' bath orbitals are specifically constructed to numerically minimize the hybridization error estimate.
\vii{
Finally, while the spectral orbitals were constructed to reproduce the real axis Green's~function, since the analytic continuation from real to imaginary axis is well conditioned, they should also be able to describe the Matsubara Green's~function in a systematically improvable fashion.
For this reason, we also compare to the spectral orbitals in Fig.~\ref{fig:porphyrin-msb}.
}

\vi{
The results of these three approaches are shown in Fig.~\ref{fig:porphyrin-msb}.
}
The Matsubara orbitals outperform the direct discretization method significantly, with only ten orbitals giving a more accurate representation of the system Green's function than 60 states from the direct discretization.
Additionally,
\vi{the projection approach}
\vii{they}
also performs slightly better than the numerical fit for the the Green's function accuracy, while at the same time being less computationally demanding, extendable beyond a few tens of orbitals and not dependent on starting parameters as compared to the numerical fitting. The numerical fit gives the best description of the hybridization for small numbers of bath orbitals, but this is to be expected as this quantity is directly minimized.
%
We note that the numerical fit could also be performed with the difference of the Green's functions, rather than hybridization in the distance functional of Eq.~\eqref{eq:numcostfn}.
In this case the numerical method should likely be the most accurate, as it explicitly minimizes this error norm\cite{Liebsch_2011}.
On the other hand the fit also becomes more computationally demanding, the analytic expressions for the gradient of the objective function more involved, and the question of appropriate starting parameters still prevails.
\vii{The spectral orbitals are far less efficient in expanding the Matsubara Green's~function than the Matsubara orbitals.
This is somewhat expected, since it is well known that Matsubara quantities can be well approximated with a small number of states, while this is in general not possible along the real frequency axis.
The Matsubara orbitals can thus make use of this flexibility, while the spectral orbitals also have to fulfill the more restrictive constraint of representing an accurate real frequency spectrum.
}
%

\vii{
\subsubsection{Bath convergence in the presence of local interactions}
%
%
To address the question of how spectral and Matsubara orbitals perform in the presence of explicit electron correlation, we apply coupled-cluster theory with single and double excitations (CCSD) to solve for the ground state of the correlated quantum cluster problem, defined by the 3d states of the iron porphyrin molecule (the system) and a set of bath orbitals describing the molecular environment, and calculate the number of 3d electrons via the one-particle reduced density matrix.
Questions regarding the occupation of orbital spaces are relevant in many applications and have been been studied for similar systems before \cite{allerdt2020many}.
We note that only the one-electron Hamiltonian is projected into the bath space, whereas an explicit unscreened electron-electron interaction term is only added for the 3d electrons.
This is equivalent to the situation in the \textit{non-interacting bath} variant of DMET \cite{wouters2016practical} and in line with DMFT, which also does not account for explicit electron interactions between impurity and bath in its standard formulation.
Mean-field interactions between system and bath electrons are still taken into account,
since the system's one-electron Hamiltonian is taken to be the Fock matrix (plus a double counting correction, to remove the mean-field interaction between 3d electrons).
The number of electrons of the cluster problem was chosen such that the total number of 3d electrons in the mean-field calculation of the cluster problem is best matched to the mean-field number from the full system calculation (approximately $6.30$).

For all kinds of bath orbitals (direct discretization, spectral orbitals, and Matsubara orbitals) and depending on the number of bath orbitals, we find two different mean-field solutions in the cluster, one which approaches an total occupation of the 3d space with approximately $6.26$ electrons with increasing bath size, and the other approaching the correct value of $6.30$.
Furthermore, the first solution is clearly unphysical, as it has different occupations of the $\mathrm{3d_{xz}}$ and $\mathrm{3d_{yz}}$ orbitals and thus breaks the point group symmetry of the molecule.
We ignore these unphysical solutions and perform the correlated CCSD method only for the mean-field solutions of the second type, which have the correct point group symmetry.
Figure~\ref{fig:porphyrin-ccsd} shows the occupation of the 3d states for different number of bath states.
\begin{figure}[!ht]
    \centering
    \includegraphics[width=\linewidth]{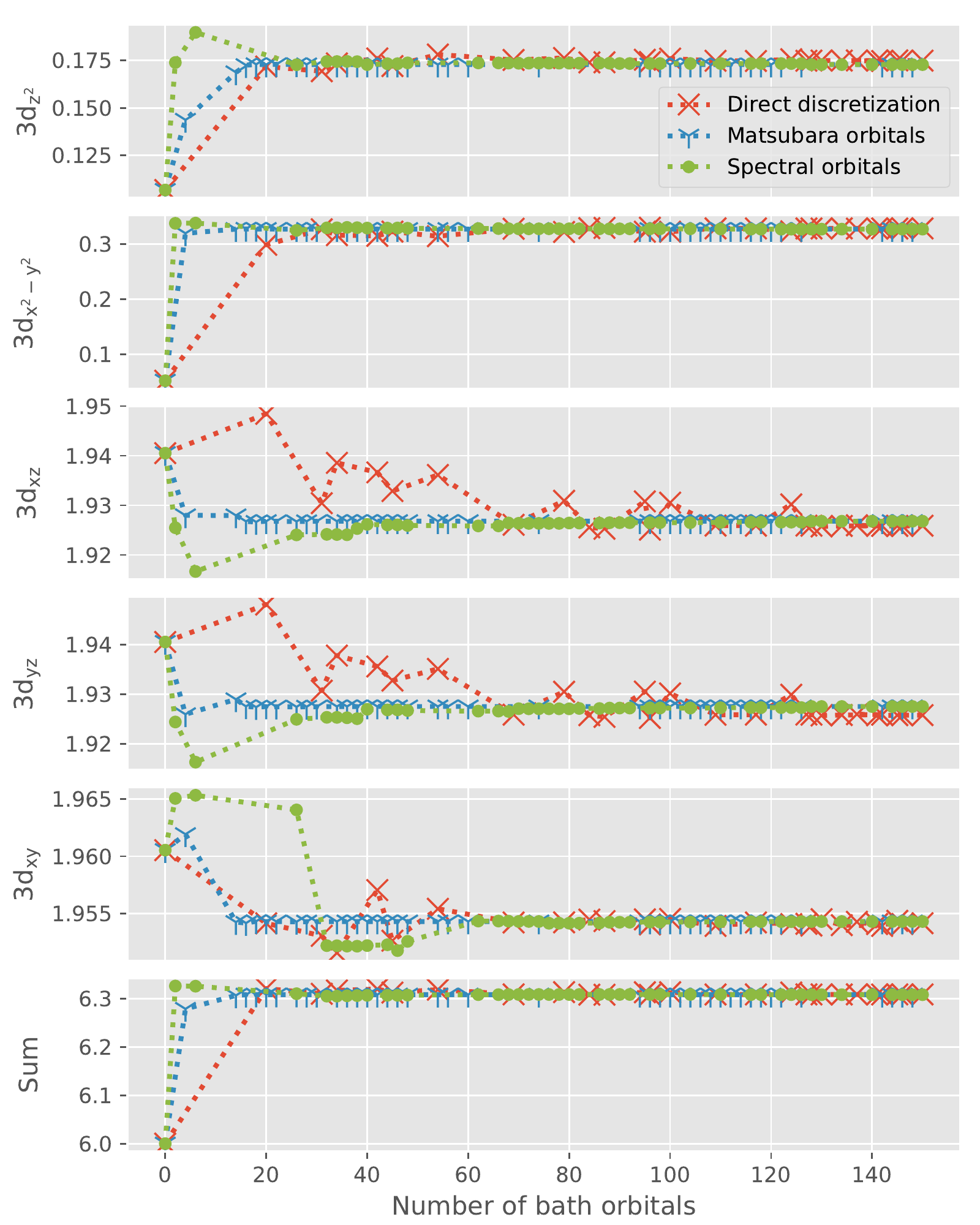}
    \caption{Occupation of iron 3d orbitals of iron porphyrin, when solved as a quantum cluster problem with CCSD for different number of bath orbitals.
    Gaps between the data points are due to the neglect of the unphysical solution, discussed in the main text.
    The four nitrogen ligands are placed on top of the $x$ and $y$ axis, with the iron atom being at the origin.}
    \label{fig:porphyrin-ccsd}
\end{figure}
The spectral and especially the Matsubara bath orbitals converge the occupation much more quickly than the direct discretization method.
Additionally, the direct discretization is less stable and more often leads to the unphysical mean-field solution, in particular when using a small number of bath orbitals.
We also note that the direct discretization has slightly higher occupation of the $\mathrm{3d_{z^2}}$ orbital and slightly smaller occupations of the $\mathrm{3d_{xz}}$ and $\mathrm{3d_{yz}}$ orbitals even when using as many as 150 bath states.
This is most likely since the hybridization is only discretized in the energy interval between $-10~E_\mathrm{h}$ and $10~E_\mathrm{h}$ and coupling to the environment outside of this energy range will not be achieved, no matter how many discretization points are used.
On the contrary, the bath orbitals derived from the projection method will always capture all energy scales of the environment \textit{eventually}, even if the Green's~function is only matched in a restricted energy interval, as is the case here.
This is because the bath orbitals are explicitly linear combination of environment states and required to be orthogonal to each other.
Overall, 20 Matsubara orbitals or 60 spectral orbitals are enough
to converge the occupation of the 3d states of iron porphyrin in the presence of local interactions.
The inclusion of the explicit interactions in the system Hamiltonian leads to a significant rearrangement of electron numbers between the correlated orbitals, with (for example) the occupation of the $3\mathrm{d}_{z^2}$ orbital increasing compared to the mean-field state from 0.05 electrons to 0.18 electrons.
Finally, it should be also noted that in the projective approach the convergence of the occupation numbers could likely be improved further, by adding the 0-th power orbitals, which ensure an exact matching of the non-interacting density matrix.
%
In this way, the remaining spectral or Matsubara orbitals only need to account for the dynamical character of the hybridization {\em beyond} that required to match the density matrix.
}


\subsection{Valence Orbitals of Benzene Coupled to Constant Hybridization}
\begin{figure}[!h]
    \centering
    \includegraphics[width=\linewidth]{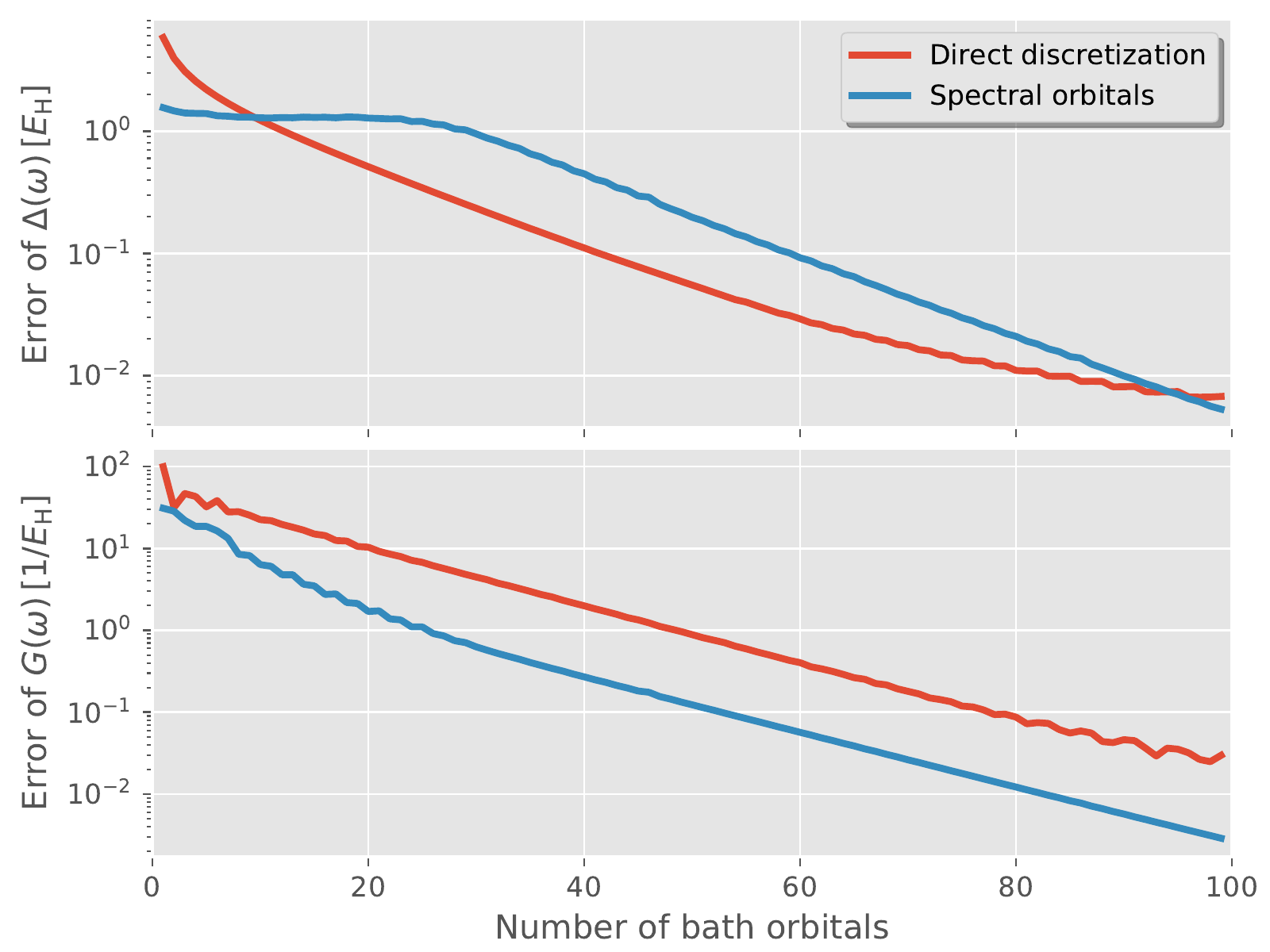}
    \caption{
    Error of the retarded (real-frequency) hybridization (top row) and Green's function (bottom row)
    defined according to Eq.~\eqref{eq:error-ret}, for the benzene system coupled to constant, continuous hybridization.
    The error is measured over 1000 uniform frequency points in the interval $[-5~\eh, 5~\eh]$ and with $\eta = 0.05~\eh$ 
    \label{fig:benzene-ret}
    }
\end{figure}

As our second test system we consider the six molecular orbitals constituting the $\pi$-system of a benzene molecule (calculated with RHF/cc-pVTZ) coupled to a constant and continuous hybridization in the energy interval $[-2~\eh,2~\eh]$ and with coupling strength $J_{ab}(\omega) = 0.1/ \pi~\eh$, which we model using 2000 uniformly distributed states in this energy range.
This `box' hybridization can model the presence of a conducting electrode in contact with the benzene molecule, for example in a molecular junction.
This simple form of the coupling matrix $\mathbf{J}(\omega)$ with equal entries everywhere means that the eigendecomposition of Eq.~\eqref{eq:direct_diag} only leads
to a single nonzero coupling vector in each frequency interval. 
We can thus expect that the direct discretization method perform especially well in this particular example for a given number of bath orbitals, as only one orbital per chosen frequency point is required.
Nonetheless, even for this system we find that the spectral orbitals outperform the direct method in their ability to efficiently approximate the real-frequency (retarded) Green's~function, as shown in Fig.~\ref{fig:benzene-ret}.
For the spectral orbitals we used 1000 frequency points in the interval $[-5~\eh, 5~\eh]$ with a broadening $\Lambda=0.05~\eh$ and the direct discretization method is performed in the same way as for the iron prophyrin system, within the support of the hybridization $[-2~\eh, 2~\eh]$.
The difference between the two methods is smaller than in the porphyrin test case, but 20 spectral orbitals are still able to represent the Green's function to a similar level of accuracy as 40 directly discretized states. Once again, the spectral orbital methods sacrifice a faithful representation of the hybridization, for an improved description of the system Green's function compared to the direct discretization of the hybridization.

\begin{figure}[!h]
    \centering
    \includegraphics[width=\linewidth]{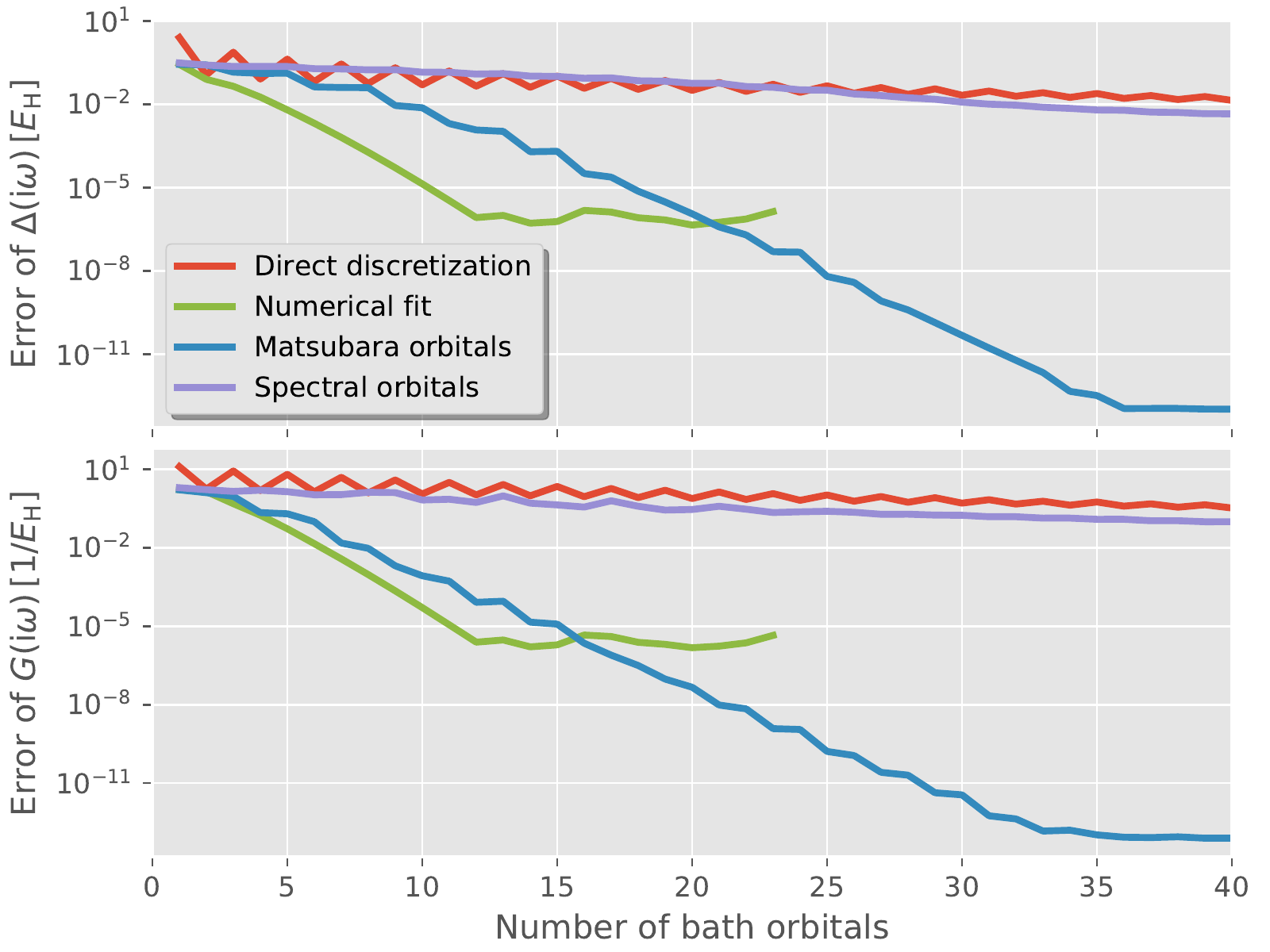}
    \caption{
    Error of the Matsubara hybridization (top row) and Green's function (bottom row)
    defined according to Eq.~\eqref{eq:error-msb} for the benzene system.
    The error is measured using the first 1000 Matsubara points with $\beta=100$.
    \label{fig:benzene-msb}
    }
\end{figure}

\vi{
We also test the Matsubara orbitals and compare them to the direct discretization of the hybridization and the numerical fitting in Fig.~\ref{fig:benzene-msb}.}
\vii{
We also test the Matsubara orbitals and compare them to the direct discretization of the hybridization, the numerical fitting, and the spectral orbitals in Fig.~\ref{fig:benzene-msb}.
}
Again the Matsubara orbitals are much better than the direct discretization and a little worse than the numerical fit in their ability to reproduce the true hybridization. However, they are the optimal choice in minimizing the Matsubara Green's~function error for all numbers of representative bath orbitals.
Furthermore, the numerical fit does not improve anymore beyond 12 bath states, most likely since with increasing number of fit parameters the minimum of the functional~\eqref{eq:numcostfn} becomes very shallow and we terminate the minimization when the norm of the gradient of the objective function becomes less than $10^{-12}$.
This threshold is quite small compared to the precision of double floating point numbers and can already lead to more than 100,000 iterations of the fitting procedure,
so that a much smaller threshold is not feasible, limiting the ultimate accuracy.
This illustrates some of the difficulties and computational expense encountered with numerical fitting approaches, which are absent in the projection method.
\vii{As seen before for the iron porphyrin, the spectral orbitals are not as efficient as Matsubara orbitals when measuring the error of Matsubara quantities and in this case barely improve upon a direct discretization method.
}

\begin{figure*}[!htb]
\subfloat[Hybridization\label{fig:benzene-spec-h}]{
    \includegraphics[width=0.48\linewidth]{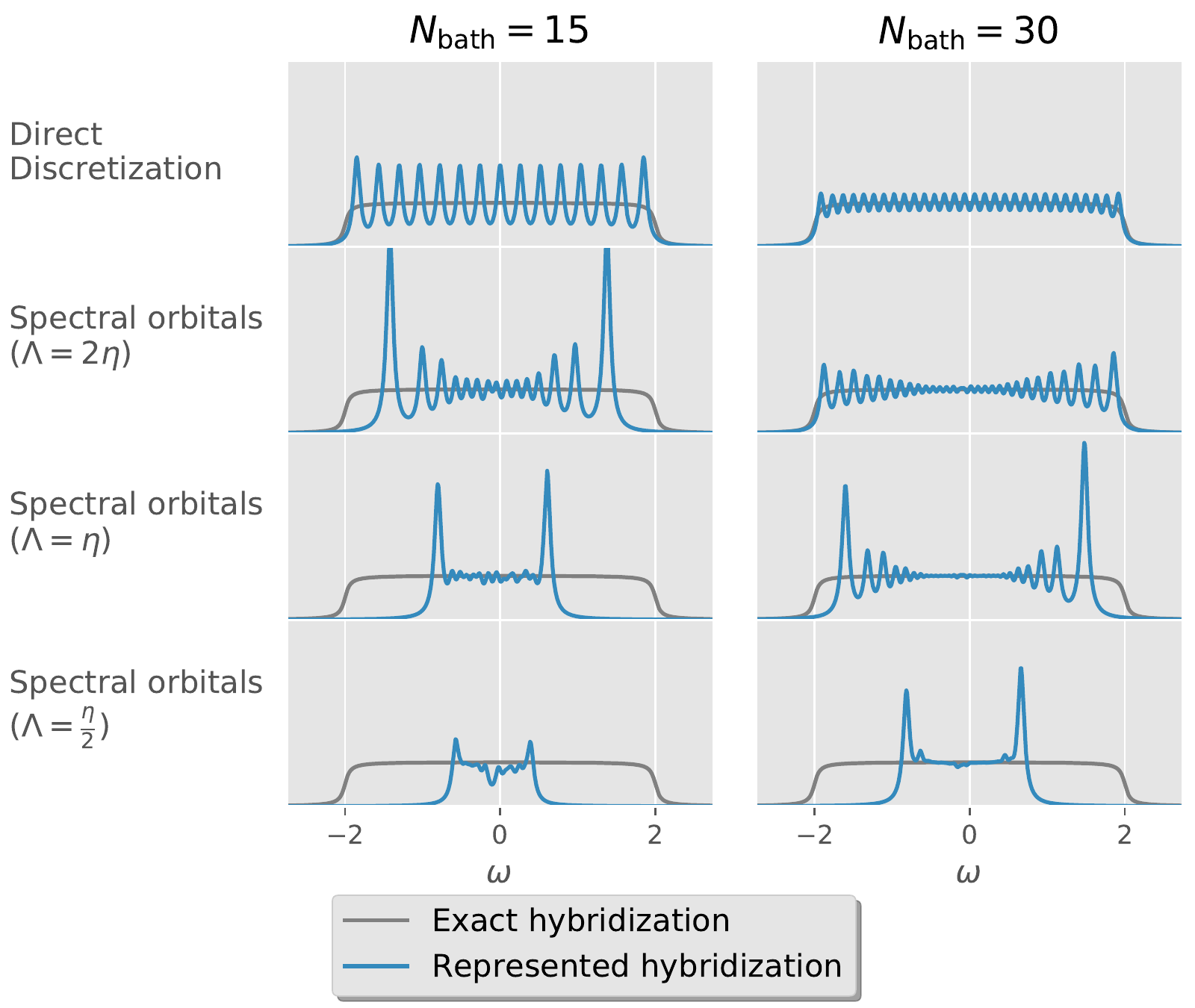}
    }
\subfloat[System Green's~function\label{fig:benzene-spec-g}]{
    \includegraphics[width=0.48\linewidth]{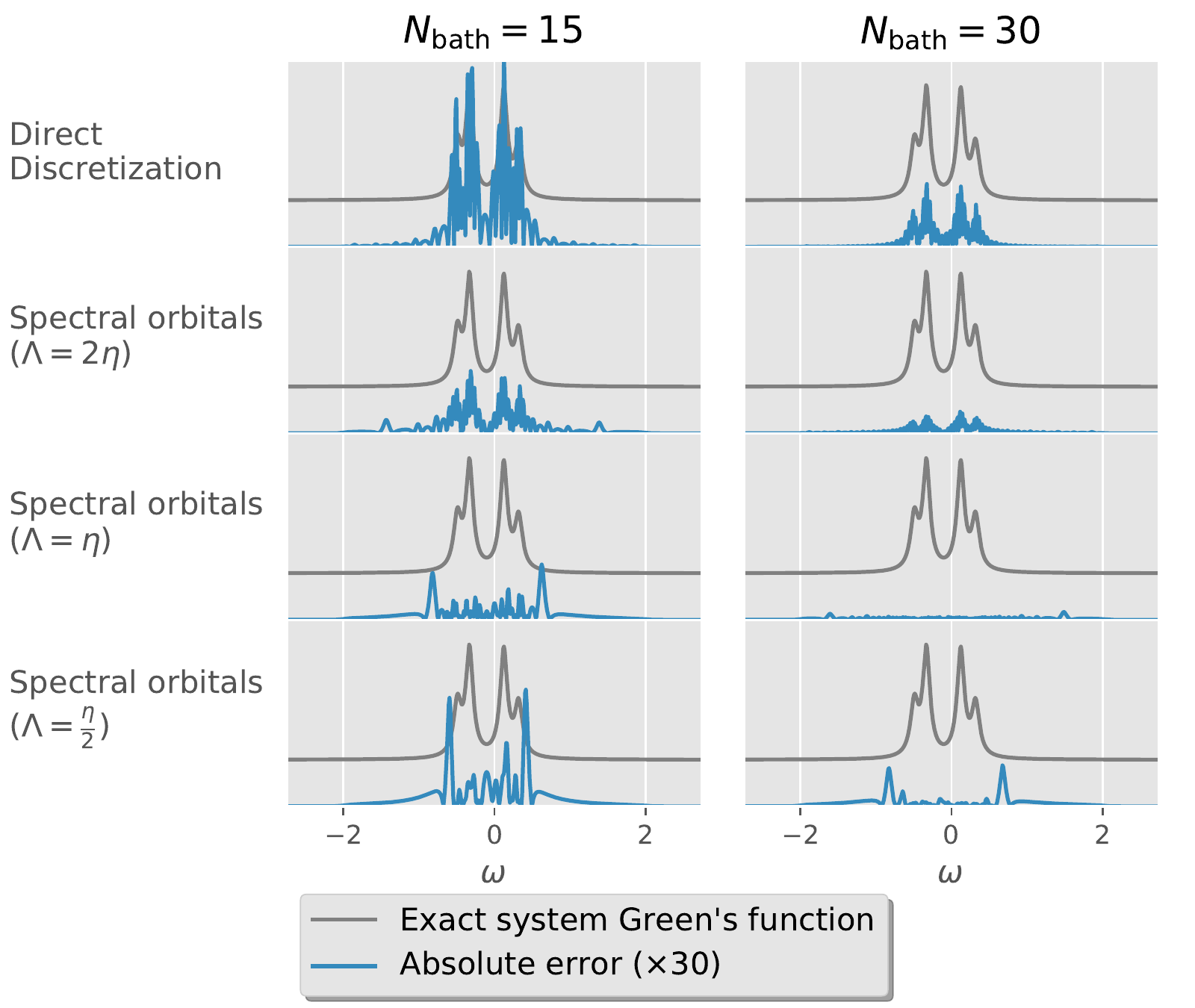}
}
    \caption{
    (a) Represented environmental spectral distributions resulting from the direct discretization method (first row) and spectral bath orbital method with different parameter $\Lambda$ (second to fourth row) for the benzene system coupled to the continuous hybridization.
    The number of bath orbitals is 15 in the first column and 30 in the second column for all methods.
    The grey line represents the exact hybridization.
    (b) Same as (a) but showing the absolute error of the imaginary part of the corresponding system Green's function for the different approaches to constructing these compressed bath orbitals.
    As an illustration, we also show the value of the imaginary part of the exact system Green's~function in the presence of the entire hybridization (shifted grey line).
    Note that the absolute errors in blue are scaled by a factor of 30 compared to the values in grey.
    A broadening of $\eta = 0.05~\eh$ was used for all plots.
    \label{fig:benzene-spec}
    }
\end{figure*}
In order to better understand the efficiency of the projection method in this case of a continuous hybridization, in Fig.~\ref{fig:benzene-spec-h}
we compare the environment spectral distribution represented by the chosen bath orbitals in the direct discretization method, to the spectral bath orbitals for 15 and 30 chosen states.
Furthermore, we also show the value and absolute error (scaled by a factor of 30) of the imaginary part of the corresponding system Green's~function in Fig.~\ref{fig:benzene-spec-g}.
Due to the equal spectral weight criterion, the direct discretization method places the energies of the bath orbitals [poles of the represented $\Delta_{\rm disc}(\omega)$] uniformly within the support of the hybridization $[-2~\eh, 2~\eh]$, which in this case is the unambiguous optimal way to represent this constant `box' hybridization, as shown in the first row of Fig.~\ref{fig:benzene-spec-h}.

However, we are primarily interested in the effect of this coupling on the system propagator, where the discrete orbital energies of the Benzene are broadened, and finite lifetimes introduced to these excitations. The direct discretization in the first row of Fig.~\ref{fig:benzene-spec-g} shows that this approach leads large errors in the system Green's~function, especially in the areas where the latter has significant spectral weight.
On the other hand, the spectral orbitals resolve this energy interval relevant to the system propagator between $\omega = -0.6~\eh$ and $\omega = 0.6~\eh$ more accurately, as seen in the represented environmental spectrum, meaning that for the same number of bath orbitals, the remaining (in this case high energy) parts are more coarsely represented, with only a smaller number of bath states.
This demonstrates that the projection method, in contrast to a direct discretization of the hybridization, takes the system's Hamiltonian into account when determining the the bath states.
The balance between the description of the regions close and far away from the poles of the system Green's function can be adjusted with the parameter $\Lambda$ used in Eq.~\eqref{eq:spectral_orbitals}, as can be seen by comparing the second to fourth row of Fig.~\ref{fig:benzene-spec-h}.
A smaller $\Lambda$ will improve the the description of the hybridization in the energetic vicinity of the Green's function while sacrificing the accuracy in the representation of the remaining energy range and vice versa.
Figure~\ref{fig:benzene-spec-g} shows that using $\Lambda = 2 \eta = 0.1~\eh$  leads to larger errors around the maxima of the Green's function, increasingly similar to the direct discretization method, whereas $\Lambda= \eta/2 = 0.025~\eh$ results in small errors around these maxima, but larger errors in other parts of the frequency range.
Finally, $\Lambda = \eta = 0.05~\eh$ leads to small errors and a balanced description of the Green's function over the whole frequency axis.
The fact that a value of $\Lambda$ equal to the broadening $\eta$ with which the error is measured (in Fig.~\ref{fig:benzene-ret}) or plotted (in Fig.~\ref{fig:benzene-spec-g}) yields the best results is not a coincidence, since the spectral bath orbitals were constructed to best represent the Green's function with this broadening in the first place.
The question of which broadening $\Lambda$ one should use in practice when constructing spectral bath orbitals thus becomes a question of the desired resolution in the spectral features of the retarded Green's function.
This question cannot be answered in a general manner but instead the answer depends on the quantities of interest, the involved energy scales, and also the maximum number of bath states the application can afford to include, as a Green's function with smaller broadening requires more states for an accurate representation.
These considerations also extend to the Matsubara orbitals, where one has a choice of the inverse temperature $\beta$, as well as the number of points (which determines a weighting between low and higher energy description) of the Green's function that should be reproduced.
The projection method, however, provides an easy to follow recipe after this specific choice has been made.
We note that the insights gained here from the analysis of the projection method could be used to improve the description of the system Green's function via the direct discretization method as well.
For example, information of the system Hamiltonian could be included in the direct method if the intervals are not selected by equal spectral weight of the hybridization, but instead by equal weight of (for example) the Green's~function, or the geometric mean of hybridization and Green's~function.
In this way, the direct method would also sample energy regions more finely where the Green's~function is significant, rather than just the hybridization.

\section{Conclusion}
In this work we presented an efficient and systematically improvable way to represent the environment of an open quantum system in terms of a compact set of bath states, such that the real frequency or Matsubara Green's functions are well approximated. This is in particular an important step within dynamical mean-field theory, where Hamiltonian-based impurity solvers are required to operate.
We call this approach ``projection method'' as it explicitly constructs bath orbitals and projects the full system $\oplus$ environment Hamiltonian into the system $\oplus$ bath space.
For the bath orbitals, we use the imaginary part of the dynamic bath orbitals of Refs.~\onlinecite{PhysRevB.91.155107,PhysRevB.101.045126},
which exactly reproduce the imaginary part the Green's function on a selected set of complex frequencies points.
The orbital space is then truncated such that only the most significant linear combinations of the dynamic bath orbitals are used.

In contrast to existing approaches like the direct discretization of the hybridization function or a numerical fit of the latter, the projection method takes
the system Hamiltonian implicitly into account and is thus able to discretize the hybridization only ``where it matter'', i.e., close to the poles of the Green's function.
We test this approach on two different systems: the 3d atomic shell of the iron atom in an iron porphyrin molecule and the $\pi$-molecular orbital system of a benzene molecule coupled to a continuous, model hybridization.
In both cases we find that the projection method leads to a more efficient description of the retarded and Matsubara Green's~functions than the direct discretization method.
Furthermore, it offers similar levels of accuracy in the description of the Matsubara Green's~function as a numerical fit, without the computational cost and practical difficulties associated with the latter.
\vii{
At the example of the iron 3d shell in iron porphyrin, we found that the projective method also performs well in the presence of local electron interactions. 
}
It therefore has potential applications when an efficient representation of a quantum environment is desired, as in quantum embedding methods or simulations of molecular scale electronics.

\section*{Acknowledgements}
G.H.B. gratefully acknowledges support from the Royal Society via a University Research Fellowship, in addition to funding from the European Union's Horizon 2020 research and innovation programme under grant agreement No. 759063. We are also grateful to the UK Materials and Molecular Modelling Hub for computational resources, which is partially funded by EPSRC (EP/P020194/1).

\section{Appendix}
\vii{
\subsection{A proposed DMFT algorithm with projective bath orbitals}\label{ssec:dmft}
The concepts outlined in the compression of an environment in this work can be applied to many settings, however an obvious and immediate approach would be for the construction of bath orbitals within a self-consistent DMFT method. Indeed, a particular choice of (frequency-dependent) projective bath orbitals was already used in a variant of DMFT by the authors in an approach denoted $\omega$-DMFT \cite{PhysRevB.101.045126}.
When applying this projective method for the bath construction of DMFT in practice,
one is faced with the problem that independent-particle environment states of Eq.~\eqref{eq:env} and their couplings in Eq.~\eqref{eq:ham_coup} are only explicitly known in the absence of any self-energy in the environment, which is only the case in the first DMFT iteration.
While it is always possible to describe the self-energy effect in the environment in terms of an auxiliary set of states, this representation of the self-energy is usually not known explicitly.
Instead, most quantum impurity solvers yield the correlated Green's~function or self-energy expressed on a frequency or time grid.
In $\omega$-DMFT, we addressed this problem by fitting the Matsubara self-energy numerically.
It might seem as if nothing has been gained by the projection method for bath orbitals in this case, since a fitting step of a dynamical quantity is still necessary. However there are two distinct advantages of this approach which remain:
\begin{enumerate}
    \item At least part of the hybridization, namely due to the one-electron Hamiltonian, will be exact, in other words the fitting error is only affecting a smaller quantity. In fact, in the absence of any electron correlation (``$U=0$'') no fitting has to be done at all, since the self-energy is zero, whereas in standard DMFT, the hybridization would still have to be fitted.
    It follows that at small interactions $U$, any error arising from the fitting will also be small, since the self-energy as a quantity is small.
    \item When fitting the hybridization in standard DMFT, the resulting bath states are used in the correlated quantum impurity problem and thus only a few can be fitted, due to the exponential scaling of the impurity solver with system size.
    When fitting the self-energy in $\omega$-DMFT, the resulting auxiliary states are only used to construct an intermediate quadratic Hamiltonian, which can be diagonalized at $\mathcal{O}(N^3)$ cost, and thus many more auxiliary states can be used.
    With this increased number of parameters, the fitting step becomes less problematic, as the penalty functional can be minimized to a lower value and local minima occur less often in higher dimensional parameter spaces.
\end{enumerate}
Finally, we note that the direct discretization method in conjunction with the projection method presented in this paper can be used to entirely circumvent any numerical fitting of dynamical quantities, if the self-energy is obtained on the real frequency axis, or indeed some representation which can be recast as a Lehmann representation on the real-frequency axis, such as the spectral moments.
%
%
%
In this way, the numerical fitting in DMFT would be avoided entirely, and is a direction we are actively pursuing. 

\subsection{Other choices of functions for projective bath orbitals}\label{ssec:legendre}
The projective approach presented in Sec.~\ref{sec:projection} allows constructing bath orbitals which can reproduce various local projectsion of full-system quantities via the choice of the function $f$ in Eqs.~(\ref{eq:bath-general-occ},\ref{eq:bath-general-vir}).
It would be particularly interesting to match a representation of the imaginary time Green's~function in terms of orthogonal polynomials, such as the Legendre polynomials, as it is well known that such representations are very compact \cite{PhysRevB.84.075145,doi:10.1021/acs.jctc.5b00884,doi:10.1021/acs.jctc.5b00884,PhysRevB.98.075127,dong2020legendre,shinaoka2017compressing}.
To construct bath orbitals which span the space which can fully describe the Green's function in the Legendre representation of a given degree $l$, one can simply use
\begin{equation}
    f(\lambda_i;l) = \int_{-\beta}^0 P_l[x(\tau)] G_{\lambda_i}(\tau) \: \mathrm{d} \tau
    ,
\end{equation}
where $P_l$ is the Legendre~polynomial of degree $l$, $x(\tau) = 2\tau / \beta + 1$
maps the interval of integration~$[-\beta, 0]$ to the interval~$[-1, 1]$, and
$G_{\lambda_i}(\tau)$ is the imaginary time Green's function
\begin{equation}
    G_{\lambda_i}(\tau) =
    \frac{\mathrm{e}^{-\lambda_i \tau}}{1 + \mathrm{e}^{\lambda_i \beta}}
    .
\end{equation}
Including these bath orbitals up to an order $l_\mathrm{max}$ results in an approximate representation of the imaginary time and frequency dynamics.
In an initial investigation we found that the resulting ``Legendre bath orbitals'' for small values of $l_\mathrm{max} < 100$) are more efficient than the Matsubara orbitals constructed using $l_\mathrm{max} + 1$ Matsubara points $\omega_n$ in Eq.~\eqref{eq:matsubara_orbitals} (such that the total number of bath orbitals are equal).
However, in this paper we follow a different approach and add a very large number of Matsubara points, usually 1000, to form the outer product matrix $P$ of Eq.~\eqref{eq:outer_product_matrix} and then perform the truncation to a smaller number after diagonalization in the eigenbasis of $P$.
The error in the Matsubara representation before truncation of $P$ is thus almost negligible and the efficiency of the Legendre representation for small values of $l_\mathrm{max}$ does not yield any benefits over using regular Matsubara frequency points.
It is also interesting to compare the Legendre bath orbitals, truncated to a small number $l_\mathrm{max}+1$ \textit{before} orthogonalization, with the Matsubara bath orbitals obtained with a large number of Matsubara points and truncated to the same number of bath orbitals \textit{after} orthogonalization.
In this case, we find that the Matsubara bath orbitals are still more compact than the Legendre bath orbitals.
We attribute this to the fact that selecting bath orbitals in an orthogonal basis ensures that no redundant space is spanned.
On the other hand, the non-orthogonalized Legendre bath orbitals will in general not be linearly independent.
Increasing $l_\mathrm{max}$ thus increasingly lead to spanning redundant space, which means that the orbitals will obtain a large arbitrary component as a result of the orthogonalization.
}

%



\end{document}